\documentclass[twocolumn,showpacs,preprintnumbers,amsmath,amssymb,floatfix,prb,showkeys]{revtex4}
\usepackage{graphicx}
\usepackage{epsfig}
\usepackage{dcolumn}
\usepackage{bm}
\usepackage{times}
\def\Xint#1{\mathchoice
   {\XXint\displaystyle\textstyle{#1}}%
   {\XXint\textstyle\scriptstyle{#1}}%
   {\XXint\scriptstyle\scriptscriptstyle{#1}}%
   {\XXint\scriptscriptstyle\scriptscriptstyle{#1}}%
   \!\int}
\def\XXint#1#2#3{{\setbox0=\hbox{$#1{#2#3}{\int}$}
     \vcenter{\hbox{$#2#3$}}\kern-.5\wd0}}

\def\dashint{\Xint-}
\begin{document}


\title{Phonon Decoherence of a Double Quantum Dot Charge Qubit}

\author{Serguei Vorojtsov,$^1$ Eduardo R. Mucciolo,$^{2,3}$ and 
Harold U. Baranger$^1$}

\affiliation{$^1$Department of Physics, Duke University, Box 90305,
Durham, North Carolina 27708-0305\\ $^2$Department of Physics,
University of Central Florida, Box 162385, Orlando, Florida
32816-2385\\ $^3$Departamento de F\a'{\i}sica, Pontif\a'{\i}cia
Universidade Cat\'olica do Rio de Janeiro, C.P. 37801, 22452-970 Rio
de Janeiro, Brazil}

\date{\today}

\begin{abstract}
We study decoherence of a quantum dot charge qubit due to coupling to
piezoelectric acoustic phonons in the Born-Markov approximation.
After including appropriate form factors, we find that phonon
decoherence rates are one to two orders of magnitude weaker than was
previously predicted.
We calculate the dependence of the $Q$-factor on lattice temperature,
quantum dot size, and interdot coupling. Our results suggest that
mechanisms other than phonon decoherence play a more significant role
in current experimental setups.
\end{abstract}




\maketitle


\section{Introduction}

Since the discovery that quantum algorithms can solve certain
computational problems much more efficiently than classical
ones,\cite{nielsen00} attention has been devoted to the physical
implementation of quantum computation. Among the many proposals, 
there are those based on the electron spin\cite{loss98,divincenzo00} 
or charge\cite{blick00,tanamoto00,fedichkin04,brandes02,wu04} in 
laterally confined quantum dots, which may have great potential for 
scalability and integration within current technologies.

Single qubit operations involving the spin of an electron in a quantum
dot will likely require precise engineering of the underlying material
or control over local magnetic fields;\cite{divincenzo99} both have
yet to be achieved in practice. In contrast, single qubit operations
involving charge in a double quantum dot (DQD) \cite{vanderwiel03} are
already within experimental reach.\cite{hayashi03,petta04} They can be
performed either by sending electrical pulses to modulate the
potential barrier between the dots (tunnel pulsing)\cite{fedichkin04,wu04}
or by changing the relative position of the energy levels (bias
pulsing).\cite{hayashi03} In both cases one acts on the overlap
between the electronic wave functions of the dots. This permits direct
control over the two low-energy charge states of the system -- the
basis states $|1\rangle$ and $|2\rangle$ of a qubit: Calling $N_{1}$
($N_{2}$) the number of excess electrons in the left (right) dot, we
have that $|1\rangle = (1,0)$ and $|2\rangle = (0,1)$.

\begin{figure}[b]
\includegraphics[width=4cm]{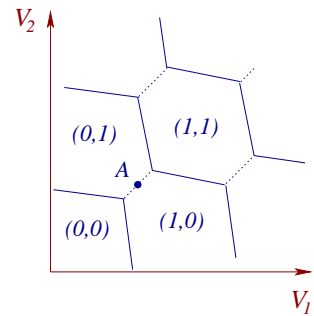}
\caption{Schematic Coulomb blockade stability diagram for a double
quantum dot system at zero bias.\cite{vanderwiel03} $(N_{1},N_{2})$
denotes the number of excess electrons in the dots for given values of
the gate voltages $V_1$ and $V_2$. The solid lines indicate
transitions in the total charge, while the dotted lines indicate
transitions where charge only moves between dots. The point $A$ marks
the qubit working point.}
\label{fig:diamonds}
\end{figure}

The proposed DQD charge qubit relies on having two lateral quantum
dots tuned to the $(1,0)\!\leftrightarrow\! (0,1)$ transition line of
the Coulomb blockade stability diagram (see Fig.~\ref{fig:diamonds}).
Along this line, an electron can move between the dots with no
charging energy cost. An advantage of this system is that the Hilbert
space is two-dimensional, even at moderate temperatures, since
single-particle excitations do not alter the charge configuration.
Leakage from the computational space involves energies of order the
charging energy which is quite large in practice ($\sim\! 1$~meV
$\sim\! 10$~K). In the case of tunnel pulsing, working adiabatically
-- such that the inverse of the switching time is much less than the
charging energy -- assures minimal leakage. The large charging energy
implies that pulses as short as tens to hundreds of picoseconds would
be well within the adiabatic regime. However, the drawback of using
charge to build qubits is the high decoherence rates when compared to
spin. Since for any successful qubit one must be able to perform
single- and double-qubit operations much faster than the decoherence
time, a quantitative understanding of decoherence mechanisms in a DQD
is essential.

In this work, we carry out an analysis of phonon decoherence in a DQD
charge qubit. During qubit operations, the electron charge movement
induces phonon creation and annihilation, thus leading to energy
relaxation and decoherence. In order to quantify these effects, we
follow the time dependence of the system's reduced density matrix,
after tracing out the phonon bath, using the Redfield formalism in the
Born and Markov approximations.\cite{argyres64,pollard94}

Our results show that decoherence rates for this situation are one to
two orders of magnitude weaker than previously estimated. The
discrepancy arises mainly due to the use of different spectral
functions. Our model incorporates realistic geometric features which
were lacking in previous calculations. When compared to recent
experimental results, our calculations indicate that phonons are
likely not the main source of decoherence in current DQD setups.

The paper is organized as follows. In Sec.~\ref{sec:model}, we
introduce the model used to describe the DQD, discuss the coupling to
phonons, and establish the Markov formulation used to solve for the
reduced density matrix. In Sec.~\ref{sec:tunnel} we study decoherence
in a single-qubit operation, while in Sec.~\ref{sec:bias} we simulate
the bias pulsing experiment of Ref.~\onlinecite{hayashi03}. Finally,
in Sec.~\ref{sec:conclusions} we present our conclusions.

\section{Model System}
\label{sec:model}

\begin{figure}[t]
\includegraphics[width=6cm]{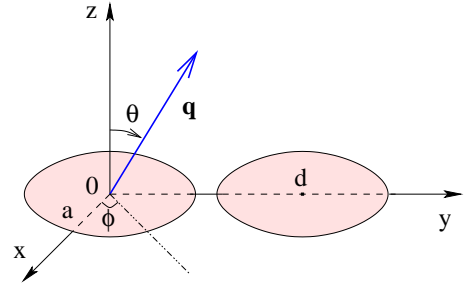}
\caption{Geometry of the double quantum dot charge qubit.}
\label{fig:geometry}
\end{figure}

We begin by assuming that the DQD is isolated from the leads. The DQD
and the phonon bath combined can then be described by the total
Hamiltonian \cite{brandes02}
\begin{equation}
\label{eq:hamilton}
H = H_{S} + H_{B} + H_{SB},
\end{equation}
where $H_S$ and $H_B$ are individual DQD and phonon Hamiltonians,
respectively, and $H_{SB}$ is the electron-phonon interaction. We
assume that gate voltages are tuned to bring the system near the
degeneracy point $A$ (Fig.~\ref{fig:diamonds}) where a single electron
may move between the two dots with little charging energy cost. To
simplify the presentation, only one quantum level on each dot is
included; $E_{1(2)}$ denotes the energy of an excess electron on the
left (right) QD (possibly including some charging energy). Likewise,
spin effects are neglected.\cite{SpinNote} Thus, in the basis $\{
\left| 1\right>$, $\left| 2 \right> \}$, the DQD Hamiltonian reads
\begin{equation}
\label{eq:hs}
H_{S} = \frac{\varepsilon (t)}{2}\, \sigma_{z} + v(t)\, \sigma_{x},
\end{equation}
where $\sigma_{z,x}$ are Pauli matrices, $\varepsilon (t) \!=\!  E_1
\!-\! E_2$ is the energy level difference, and $v(t)$ is the tunneling
amplitude connecting the dots. Notice that both $\varepsilon$ and $v$
may be time dependent. The phonon bath Hamiltonian has the usual form
($\hbar = 1$)
\begin{equation}
\label{eq:hb}
H_{B} = \sum_{\bf q} \omega_{\bf q}\, b^{\dagger}_{\bf q}b_{\bf q},
\end{equation}
where the dispersion relation $\omega_{\bf q}$ is specified below. The
electron-phonon interaction has the linear coupling
form,\cite{brandes02,brandes99}
\begin{equation}
\label{eq:hsb}
H_{SB} = \sum_{\bf q} \sum_{i=1}^2 \alpha_{\bf q} ^{(i)}\, N_i\,
\left( b_{\bf q}^\dagger + b_{-{\bf q}} \right),
\end{equation}
where $N_i$ is the number of excess electrons in the $i$-th dot and
$\alpha_{\bf q}^{(i)} = \lambda_{\bf q}\, e^{-i {\bf q} \cdot {\bf
R}_i}\, P_i ({\bf q})$, with ${\bf R}_1 \!=\! 0$ and ${\bf R}_2 \!=\!
{\bf d}$ the dot position vectors, see Fig.~\ref{fig:geometry}. The
dependence of the coupling constant $\lambda_{\bf q}$ on the material
parameters and on the wave vector ${\bf q}$ will be specified
below. The dot form factor is
\begin{equation}
P_i({\bf q}) = \int d^3r\, n_i({\bf r})\, e^{-i {\bf q} \cdot {\bf
r}},
\end{equation}
where $n_i({\bf r})$ is the excess charge density in the $i$-th dot. 
With no significant loss of generality, we will assume that the form 
factor is identical for both dots and, therefore, drop the $i$ index 
hereafter. In the basis $\{ \left| 1 \right>$, $\left| 2 \right> \}$, 
after dropping irrelevant constant terms, the electron-phonon 
interaction simplifies to
\begin{equation}
H_{SB} = K\, \Phi,
\label{eq:hsbnew}
\end{equation}
where
\begin{equation}
\label{eq:KandPhi}
K = \frac{1}{2}\, \sigma_{z} \qquad \mbox{and} \qquad
\Phi = \sum_{\bf q} g_{\bf q}\, \left( b^{\dagger}_{\bf q} + 
b_{- \bf q} \right),
\end{equation}
with $g_{\bf q} \!=\! \lambda_{\bf q}\, P({\bf q})\, \left( 1- e^{-i
{\bf q} \cdot {\bf d}} \right)$. The phonons propagate in three
dimensions, while the electrons are confined to the plane of the
underlying two-dimensional electron gas (2DEG). Notice that the
electron-pho\-non coupling is not isotropic for the DQD
(Fig.~\ref{fig:geometry}): Phonons propagating along $\phi \!=\! 0$
and any $\theta$ do not cause any relaxation, while coupling is
maximal along $\phi \!=\! \theta \!=\! \pi/2$ direction. We neglect 
any mismatch in phonon velocities at the GaAs/AlGaAs interface, 
where the 2DEG is located.
%

We now proceed with the Born-Markov-Redfield
treatment\cite{argyres64,pollard94} of this system. While the Born
approximation is clearly justified for weak electron-phonon
interaction, the Markov approximation requires, in addition, that the
bath correlation time is the smallest time scale in the problem. These
conditions are reasonably satisfied for lateral GaAs quantum dots, as
we will argue below.

Let us assume that the system and the phonon bath are disentangled
at $t=0$. Using Eqs.~(\ref{eq:hs}), (\ref{eq:hb}), and (\ref{eq:hsbnew}), 
we can write the Redfield equation for the reduced density matrix 
$\rho(t)$ of the DQD \cite{argyres64,pollard94},
\begin{equation}
\label{eq:sigmadot}
{\dot \rho (t)} = -i\left[ H_{S}(t), \rho (t)\right] +
\left\{ \left[\Lambda (t)\rho (t), K\right] + \mbox{H.c.} \right\} \;.
\end{equation}
The first term on the right-hand side yields the Liouvillian evolution
and the other terms yield the relaxation caused by the phonon
bath. The auxiliary matrix $\Lambda$ is defined as
\begin{equation}
\label{eq:defoflambda}
\Lambda (t) = \int_{0}^{\infty} d\tau\, B(\tau )\, e^{-i\tau
H_{S}(t)}\, K\, e^{i\tau H_{S}(t)}
\end{equation}
where $B(\tau ) \!=\! {\rm Tr}_{b} \{\Phi (\tau )\Phi (0) f(H_{B})\}$
is the bath correlation function, $\Phi (\tau ) \!=\! e^{iH_{B}\tau}\,
\Phi\, e^{-iH_{B}\tau}$, and $f(H_{B}) \!=\! e^{-\beta H_{B}}/{\rm
Tr}_{b}\{ e^{-\beta H_{B}} \}$, with $\beta \!=\! 1/T$ the inverse
lattice temperature ($k_B = 1$).

Using Eq.~(\ref{eq:hb}) in the definition of the bath correlation
function, we find that the latter can be expressed in the form
\begin{equation}
\label{eq:bath}
B(\tau ) = \int_{0}^{\infty} \!\! d\omega\, \nu (\omega )\, \{
e^{i\tau\omega} n_{B}(\omega ) + e^{-i\tau\omega} [1+n_{B}(\omega
)]\},
\end{equation}
where $n_{B}(\omega )$ is the Bose-Einstein distribution function and
\begin{equation}
\nu (\omega ) = \sum_{\bf q} |g_{\bf q}|^{2}\, \delta (\omega
-\omega_{\bf q})
\end{equation}
is the spectral density of the phonon bath. 

We now specialize to linear, isotropic acoustic phonons: $\omega_{\bf
q} \!=\! s |{\bf q}|$, where $s$ is the phonon velocity. Moreover, we
only consider coupling to longitudinal piezoelectric phonons,
neglecting the deformation potential contribution. For bulk GaAs, this
is justifiable at temperatures below approximately
10~$K$.\cite{bruus93} Thus,
\begin{equation}
\label{eq:piezo}
|\lambda_{\bf q}|^2 = \frac{g_{\rm ph}\, \pi^2 s^2}{\Omega |{\bf q}|}, 
\end{equation}
where $g_{\rm ph}$ is the piezoelectric constant in dimensionless form
($g_{\rm ph} \!\approx\!  0.05$ for GaAs \cite{brandes99,bruus93}) and
$\Omega$ is the unit cell volume.

The excess charge distribution in the dots is assumed Gaussian:
\begin{equation}
\label{eq:chargedensity}
n({\bf r}) = \delta (z)\, \frac{1}{2\pi a^{2}}\exp \left(
-\frac{x^2+y^2}{2a^2}\right).
\end{equation}
This is certainly a good approximation for small dots with few
electrons, but becomes less accurate for large dots. The resulting
form factor reads
\begin{equation}
\label{eq:formfactor}
P({\bf q}) = e^{-(q_x^2 + q_y^2)a^2/2}.
\end{equation}
Note that this expression differs from that in Refs.~\onlinecite{fedichkin04}
and \onlinecite{wu04} where a three-dimensional Gaussian charge
density was assumed.

Using Eqs.~(\ref{eq:piezo}) and (\ref{eq:formfactor}), as well as the
DQD geometry of Fig.~\ref{fig:geometry}, we get
\begin{eqnarray}
\label{eq:specdensity}
\nu (\omega ) = g_{\rm ph} \omega \int_{0}^{\pi /2} d\theta
\sin\theta\exp \left(
-\frac{\omega^{2}a^{2}}{s^{2}}\sin^{2}\theta\right) \\ \nonumber
\times \left[ 1 - J_{0}\left(\frac{\omega
d}{s}\sin\theta\right)\right].
\end{eqnarray}
It is instructive to inspect the asymptotic limits of this
equation. At low frequencies, $\nu (\omega \rightarrow 0) \!\approx\!
g_{\rm ph}\, d^2\, \omega^{3}/6s^2$; thus, the phonon bath is
superohmic. At high frequencies,
\begin{eqnarray}
\label{eq:omegainfty}
\nu (\omega \rightarrow \infty) \approx \frac{g_{\rm ph}\, s^2}
{a^2\omega} f\left(\frac{d}{a}\right),
\end{eqnarray}
where
\begin{eqnarray}
f \left( \frac{d}{a} \right) = \int_{0}^{\infty} dx~x~e^{-x^2}\left[ 1
- J_{0}\left(\frac{d}{a}x\right)\right].
\end{eqnarray}
Notice that the spectral function does not have the exponential decay
familiar from the spin-boson model, but rather falls off much more
slowly: $\nu (\omega \!\rightarrow\! \infty ) \!\propto\!
\omega^{-1}$. This should be contrasted with the phenomenological
expressions used in Ref.~\onlinecite{brandes02}. 

The characteristic frequency of the maximum in 
$\nu (\omega )$ is $\tau_c^{-1} = s/a$.
For typical experimental setups, $a \!\approx\!  50$ nm 
while $s \!\approx\!  5 \times 10^{3}$ m/s for GaAs, yielding 
$\tau_c \!\approx\!  10$ ps ($\tau_c^{-1} \!\approx\! 65\,\mu$eV). 
Thus, the Markovian approximation can be justified for time scales 
$t > \tau_c$ and if all pulse operations are kept adiabatic 
on the scale of $\tau_c$.

\section{Decay of Charge Oscillations}
\label{sec:tunnel}

One can operate this charge qubit in two different ways: (i) by
pulsing the tunneling amplitude $v(t)$ keeping $\varepsilon$ constant,
or (ii) by changing the energy level difference $\varepsilon (t)$
keeping $v$ constant (bias pulsing). Tunnel pulsing seems advantageous
as it implies fewer decoherence channels and less leakage. However, a
recent experiment used a bias pulsing scheme.\cite{hayashi03}

Our system's Hilbert space is two-dimensional by construction [see
Eq.~(\ref{eq:hs})], hence there is no leakage to states outside the
computational basis. We can, therefore, use square pulses instead of
smooth, adiabatic ones. This not only allows us to analytically solve
for the time evolution of the reduced density matrix,
Eq.~(\ref{eq:sigmadot}), but also renders our results applicable to
both tunnel and bias pulsing. Indeed, in both regimes one has
$\varepsilon (t) \!=\! 0$ and $v(t) \!=\! v_{m}$ for $t>0$, taking
that the pulse starts at $t=0$. Let us assume that the excess electron
is initially in the left dot: $\rho_{11}(0) \!=\! 1$ and $\rho_{12}(0)
\!=\! 0$. In this case, since the coefficients on the right-hand side
of (\ref{eq:sigmadot}) are all constants at $t>0$, we can solve the
Redfield equation exactly (see Appendix~\ref{sec:appendix} for details). 
As $\rho (t)$ has only three real independent components, the solution is
\begin{eqnarray}
\label{eq:sol1}
\rho_{11}(t) & = & \frac{1}{2}+\frac{1}{2} e^{-\frac{\gamma_{1}}{2}t}
(\cos\omega t +\frac{\gamma_{1}}{2\omega}\sin\omega t),
\\[0.05in]
\label{eq:sol2}
\mbox{Re}\, \rho_{12}(t) & = & -\frac{1}{2}(1-e^{-\gamma_{1}t})
\tanh\frac{v_{m}}{T},
\\[0.05in]
\label{eq:sol3}
\mbox{Im}\, \rho_{12}(t) & = & \frac{2v_{m} +
\gamma_{2}}{2\omega}\, e^{-\frac{\gamma_{1}}{2}t} \sin\omega t,
\end{eqnarray}
where
\begin{eqnarray}
\label{eq:omega}
\omega & = & \left[ 4v_{m}\left( v_{m}+\frac{\gamma_2}{2} \right)
-\frac{\gamma_1^2}{4} \right]^{1/2},
\\
\label{eq:gamma1}
\gamma_{1} & = & \frac{\pi}{2} \,\nu (2v_{m}) \coth \frac{v_{m}}{T},
\\
\label{eq:gamma2}
\gamma_{2} & = & -\,\dashint_{0}^{\infty}\frac{dy}{y^{2}-1}
\nu (2v_{m}y) \coth \frac{v_{m}y}{T}.
\end{eqnarray}
Note that $\gamma_{1,2} \!\ll\! v_{m}$.
We extract the customary energy and phase relaxation times, $T_1$ 
and $T_2$, by rotating to the energy eigenbasis
$\{ \left| - \right>$, $\left| + \right> \}$:
\begin{eqnarray}
\rho_{--}(t) & = & \frac{1}{2} - \mbox{Re}\, \rho_{12}(t),
\\
\rho_{-+}(t) & = & -\frac{1}{2} + \rho_{11}(t) + i\, \mbox{Im}\, \rho_{12}(t).
\end{eqnarray}
Then, the damping of the
oscillations in the diagonal matrix elements is the signature of
energy relaxation, while the phonon-induced decoherence is seen in the
exponential decay of the off-diagonal elements. For the DQD, we find
$T_1 \!=\! \gamma_1^{-1}$ and $T_2 \!=\! 2\gamma_1^{-1}$ for the
decoherence time.

The quality factor of the charge oscillations in Eq.~(\ref{eq:sol1})
is $Q = \omega/ \pi\gamma_{1}$. Using Eqs.~(\ref{eq:omega}),
(\ref{eq:gamma1}), and (\ref{eq:specdensity}), we find that
\nobreak\begin{widetext}
\begin{equation}
\label{eq:qfactor}
Q \approx \frac{4\tanh (v_{m}/T)} {\pi^{2}g_{\rm ph}} \,
\left\{ \int_{0}^{1} \frac{dx} {\sqrt{1-x}}\, 
e^{- (v_{m}/\omega_a)^2\, x}
\left[ 1 - J_{0} \left( \frac{d}{a} 
\frac{v_{m}}{\omega_a} \sqrt{x} \right) \right] \right\}^{-1},
\end{equation}
\end{widetext}
where $\omega_a = s/2a$. The $Q$-factor depends on the tunneling
amplitude $v_{m}$, lattice temperature $T$, dot radius $a$, and
interdot distance $d$.

\begin{figure}[b]
\includegraphics[width=7.4cm]{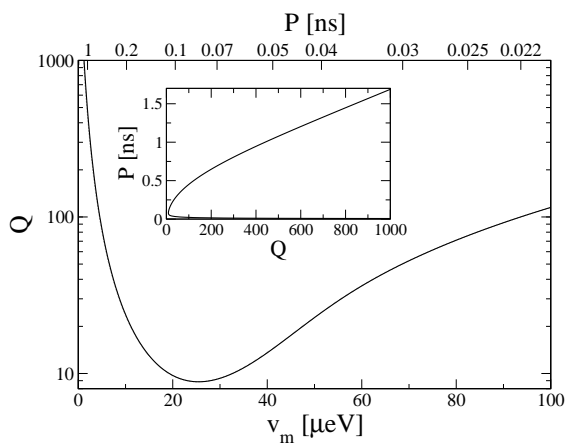}
\caption{The charge oscillation $Q$-factor as a function of the
tunneling amplitude $v_{m}$ (lower scale) and of the oscillation
period $P$ (upper scale) for a GaAs double quantum dot system. The
lattice temperature is $15$ mK and the dot radius and interdot
distance are 60 nm and 180 nm, respectively. The inset shows the
relation between $P$ and $Q$ at small tunneling amplitudes (large
periods).}
\label{fig:qfactor-Tc}
\end{figure}

Several experimental realizations of DQD systems recently appeared in
the literature.\cite{hayashi03,petta04,jeong01,chen04} In principle,
all these setups could be driven by tunnel pulsing to manipulate
charge and perform single-qubit operations. To understand how the
$Q$-factor depends on the tunneling amplitude $v_{m}$ in realistic
conditions, let us consider the DQD setup of Jeong and
coworkers.\cite{jeong01} In their device, each dot holds about 40
electrons and has a lithographic diameter of 180~nm. The effective
radius $a$ is estimated to be around 60~nm based on the device
electron density. Therefore, $d/a \!\approx\!  3$. The lattice base
temperature is 15 mK. Introducing these parameters into
Eq.~(\ref{eq:qfactor}), one can plot $Q$-factor as a function of
$v_{m}$ or, equivalently, as a function of the period of the charge
oscillations $P = 2\pi /\omega \!\approx\! \pi /v_{m}$. This is shown
in Fig.~\ref{fig:qfactor-Tc}.

To stay in the tunnel regime $v_{m}$ should be smaller than the mean
level spacing of each QD, approximately 400 $\mu$eV in the
experiment.\cite{jeong01} Therefore, in Fig.~\ref{fig:qfactor-Tc} we
only show the curve for $v_{m}$ up to 100~$\mu$eV. One has to recall
that at these values the Markov approximation used in the Redfield
formulation is not accurate (see end of Sec.~\ref{sec:model}), 
and so our results are only an estimate for $Q$. 
For strong tunneling amplitudes, when $25\, \mu{\rm
eV}<v_{m}<100\, \mu{\rm eV}$, the largest value we find for $Q$ is
close to 100. For weak tunneling with $v_{m}<25\, \mu{\rm eV}$, the
situation is more favorable and larger quality factors (thus
relatively less decoherence) can be achieved. Nevertheless, the
one-qubit operation time, which is proportional to the period, grows
linearly with $Q$ in the region of $v_{m} \to 0$, as shown in the
inset of Fig.~\ref{fig:qfactor-Tc}. Therefore, at a certain point
other decoherence mechanisms are going to impose an upper bound on
$Q$.

\begin{figure}[b]
\includegraphics[width=7.4cm]{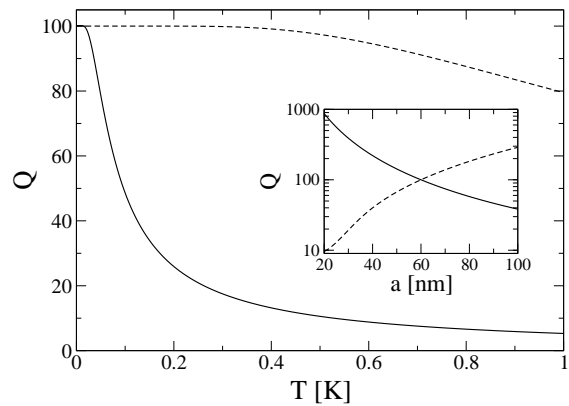}
\caption{The charge oscillation $Q$-factor as a function of the
lattice temperature. Inset: as a function of the dot radius for a
fixed ratio $d/a=3$. The solid (dashed) line corresponds to the weak
$v_{m} \!\simeq\! 53$~mK (strong $v_{m} \!\simeq\! 1.1$~K) tunneling
regime. Other parameter values are equal to those in
Fig.~\ref{fig:qfactor-Tc}.}
\label{fig:qfactor-T}
\end{figure}

The minimum of $Q$ in Fig.~\ref{fig:qfactor-Tc} occurs when $v_{m}$
coincides with the frequency at which the phonon spectral density is
maximum. It corresponds to the energy splitting between bonding and
anti-bonding states of the DQD, $2v_{m}$, being approximately equal to
the frequency of the strongest phonon mode $s/a$: $v_{m} \simeq
\omega_{a}$.

From Fig.~\ref{fig:qfactor-Tc}, it is evident that one can reach
certain values for the $Q$-factor (say, $Q = 100$) at both weak
($v_{m}\simeq 4.6~\mu$eV $\simeq 53$~mK) and strong ($v_{m}\simeq
93~\mu$eV $\simeq 1.1$~K) tunneling. However, these two regimes are
not equally convenient. From Eq.~(\ref{eq:qfactor}), it is clear that
the temperature dependence of the $Q$-factor is fully determined by
the bonding-antibonding splitting energy $2v_{m}$: $Q(T) \!=\!
Q(0)\tanh (v_{m}/T)$. We notice that $Q(T) \!\approx\!  Q(0)$ if $T\ll
v_{m}$; therefore, the $Q$-factor is less susceptible to temperature
variations for strong tunneling (Fig.~\ref{fig:qfactor-T}). Another
parameter that influences the $Q$-factor is the dot radius, which
controls the frequency of the strongest phonon mode, $s/a$. In the
strong tunneling regime (dashed curve in the inset to
Fig.~\ref{fig:qfactor-T}), one has to increase the QD size to improve
the $Q$-factor. This would reduce the energy level spacing, hence only
moderate improvement in $Q$-factor is possible. In contrast, in the
weak tunneling regime (solid curve in the inset to
Fig.~\ref{fig:qfactor-T}) one has to reduce the QD size. This can lead
to a significant (up to one order of magnitude) $Q$-factor
improvement.

\section{Bias Pulsing}
\label{sec:bias}

In a recent experiment,\cite{hayashi03} Hayashi and coworkers studied
charge oscillations in a bias-pulsed DQD. In this regime the energy
difference between the left and right-dot single-particle energy
levels is a function of time: $\varepsilon (t) = \varepsilon_0\,
u(t)$. A typical profile used for pulsing is
\begin{eqnarray}
\label{eq:biaspulse}
u(t) = 1 - \frac{1}{2} \left( \tanh\frac{t+W/2}{2\tau}
-\tanh\frac{t-W/2}{2\tau} \right),
\end{eqnarray}
where $W$ represents the pulse width and $\tau$ controls the rise and
drop times. During bias pulsing, the tunneling amplitude is kept
constant. In Ref.~\onlinecite{hayashi03}, the difference in energy
levels was induced by applying a bias voltage between left and right
leads (and not by gating the dots separately). For their setup, the
maximum level splitting amplitude was $\varepsilon_0 \!\approx\!  30\,
\mu{\rm eV}$ and $\tau \!\approx\!  15\,{\rm ps}$, corresponding to an
effective ramping time of about 100~ps.\cite{pulseprofile} The
tunneling amplitude was kept constant and estimated as $v \!\approx\!
5\,\mu{\rm eV}$, which amounts to charge oscillations with period $P
\!\approx\!  1\,{\rm ns}$.  The lattice temperature was $20\,{\rm
mK}$. Each quantum dot contained about 25 electrons and the effective
dot radius is estimated to be around $50\,{\rm nm}$ based on the
device electron density. From the electron micrograph of the device
one finds $d \!\approx\!  225\,{\rm nm}$, hence $d/a \!\approx\!
4.5$. When substituting these values into Eq.~(\ref{eq:qfactor}), one
finds $Q \!\approx\!  54$.

However, from the experimental data one observes $Q \!\approx\! 3$.
Low $Q$-factors were also obtained by Petta and coworkers in an
experiment where coherent charge oscillations in a DQD were detected
upon exciting the system with microwave radiation.\cite{petta04} Other
mechanisms of decoherence do exist in these systems, such as
background charge fluctuations\cite{itakura03} and electromagnetic
noise emerging from the gate voltages. Our results combined with the
recent experiments indicate that these other mechanisms are more
relevant than phonons.

We now turn to yet another possible source of decoherence: Leakage to
the leads when the pulse is on.\cite{hartmann04} To illustrate this
alternative source of damping of charge oscillations, we simulate the
bias-pulsing experiment of Ref.~\onlinecite{hayashi03} by implementing
a rate equation formalism similar to that used in
Ref.~\onlinecite{fujisawa04}. The formalism is based on a transport
theory put forward for the strongly biased
limit.\cite{nazarov93,gurvitz96} First, we find the stationary current
$I_0$ through the DQD structure when the pulse is off (that is, the
bias is applied):\cite{gurvitz96}
\begin{eqnarray}
\label{eq:i0}
I_{0} = e~\frac{\Gamma_L \Gamma_R}{\Gamma_L + \Gamma_R}
~\frac{v^{2}}{v^2 +\frac{\Gamma_L\Gamma_R}{4} +
\frac{\varepsilon_0^2\Gamma_L\Gamma_R}{(\Gamma_L +\Gamma_R)^2}},
\end{eqnarray}
where $e$ is the elementary charge. $\Gamma_{L(R)}$ is the partial
width of the energy level in the left (right) dot due to coupling to
the left (right) lead (when the bias is applied); in the
experiment,\cite{hayashi03} $\Gamma_{L,R} \!\approx\!  30\,\mu{\rm
eV}$. On the other hand, when the pulse is on, the stationary current
is zero. We now apply the pulse $\varepsilon (t)$ and measure the
current $I(t)$. In the experiments, the level widths $\Gamma_{L,R}$
decrease upon biasing the system. To include that effect here, we also
pulse them: $\Gamma_L(t) \!=\!  \gamma_L + (\Gamma_L - \gamma_L) u(t)$
and analogously for $\Gamma_R$, where $\gamma_{L(R)}$ is the residual
leakage to the left (right) lead when the pulse is on. We use
$\gamma_{L,R} \!=\!  0.3\,\mu{\rm eV}$, even though the real leakage
in the experiment was likely much smaller. To obtain the response
current one subtracts the stationary component: $I_{\rm resp}(t) \!=\!
I(t) - I_0 u(t)$.

\begin{figure}
\includegraphics[width=7.4cm]{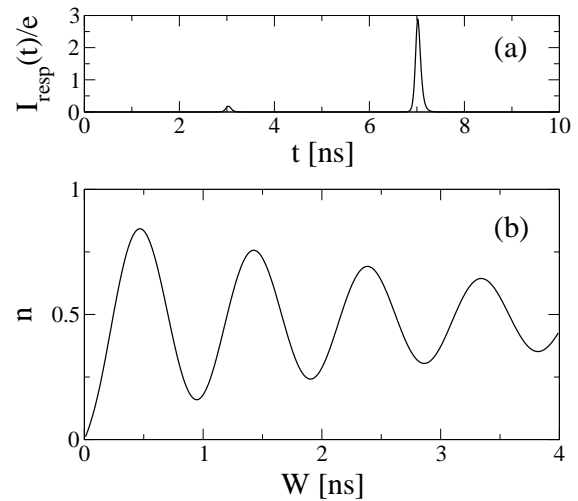}
\caption{(a) The response current $I_{\rm resp}(t)/e$ in ns$^{-1}$ as
a function of time for a pulse with $W \!=\! 4$ ns and $\tau \!=\!
30$ ps. (b) Number of electrons transfered between left and right
leads, as defined in Eq.~(\ref{eq:ncycle}), as a function of the pulse
width $W$.}
\label{fig:respcurrent}
\end{figure}

Figure \ref{fig:respcurrent}(a) shows the response current for a pulse
of width $W \!=\! 4$~ns and $\tau \!=\! 30$~ps. The latter is
approximately twice as large as in the experiment and is chosen to
enhance the effect. In Ref.~\onlinecite{hayashi03}, pulses were
applied at a frequency $f \!=\! 100$~MHz. The average number of
electrons transfered from the left to the right lead per cycle minus
that in the stationary regime is\cite{fujisawa04}
\begin{eqnarray}
\label{eq:ncycle}
n = \int_0^{1/f}\!dt\; I_{\rm resp}(t)/e.
\end{eqnarray}
(In the simulations there is no need to apply a sequence of pulses.)
Notice that $n$ oscillates as a function of the pulse width $W$ [see
Fig.~\ref{fig:respcurrent}(b)] as observed in the experiment. Two main
conclusions can be drawn from our simulation. First, the larger
$\tau$, the smaller the visibility of the charge
oscillations.\cite{fujisawa04} Second, the larger the leakage rates
$\gamma_{L,R}$ when the pulse is on, the stronger the damping of the
oscillations. While the damping due to leakage is presumably too weak
an effect to discern in the data presented in
Ref.~\onlinecite{hayashi03}, the loss of visibility due to finite $\tau$
is likely one of the causes of the small amplitude seen
experimentally.

\section{Conclusions}
\label{sec:conclusions}

The main conclusion of the paper is that, under realistic conditions,
phonon decoherence is one to two orders of magnitude weaker than
expected.\cite{brandes02,fedichkin04,wu04} The analytical expression for
the $Q$-factor given in Eq.~(\ref{eq:qfactor}) was found using an
expression for the phonon spectral density,
Eq.~(\ref{eq:specdensity}), which takes into account important
information concerning the geometry of the double quantum dot
system. In a previous work\cite{brandes02} an approximate,
phenomenological expression, $\nu (\omega )\propto\omega\exp
(-\omega /\omega_{c})$, was utilized in the treatment of charge 
qubits. There is a striking difference between these two expressions
in both the high- and low-frequency limits. Moreover, an arbitrary 
coupling constant was adopted in Ref.~\onlinecite{brandes02} 
to model the electron-phonon interaction while our treatment uses 
a value known to describe the most relevant phonon coupling in GaAs. 
On the other hand, other previous work\cite{fedichkin04,wu04}
assumed a spherically symmetric excess charge distribution 
in the dot while we have assumed a two-dimensional pancake form.
These differences account for most of the discrepancy 
between the present and previous results.

Based on these findings we conclude that phonon decoherence is too
weak to explain the damping of the charge oscillations seen in recent
experiments.\cite{hayashi03,petta04} Charge leakage to the leads
during bias pulsing is an additional source of damping, as shown in
Fig.~\ref{fig:respcurrent}(b); however, for realistic
parameters,\cite{hayashi03,fujisawa04} it turns out to be a weak
effect as well. Hence, other decoherence mechanisms, such as
background charge fluctuations or noise in the gate voltages, play the
dominant role.\cite{itakura03}

There are two distinct ways to operate a double quantum dot charge
qubit: (i) by tunnel pulsing or (ii) by bias pulsing. Tunnel pulsing
seems advantageous due to the smaller number of possible decoherence
channels. In addition, the bias pulsing scheme, in contrast to tunnel
pulsing, introduces significant loss of visibility in the charge
oscillations.

In this work we did not attempt to study leakage or loss of fidelity
due to non-adiabatic pulsing, which are both important issues for {\it
spin}-based quantum dot qubits.\cite{vorojtsov04} Moreover, we have
not attempted to go beyond the Markov approximation when deriving an
equation of motion for the reduced density matrix. Both of these
restrictions in our treatment impose some limitations on the accuracy
of our results, especially for large tunneling amplitudes.

Finally, it is worth mentioning that some extra insight would be
gained by measuring the $Q$-factor as a function of the tunneling
amplitude $v_{m}$ experimentally. Such a measurement would allow one
to map the spectral density of the boson modes responsible for the
decoherence. This would provide very valuable information about the
leading decoherence mechanisms in double quantum dot systems.

\begin{acknowledgments}
We thank A.~M.~Chang, M.~Hentschel, E.~Novais, G.~Usaj, and F.~K.~Wilhelm 
for useful discussions. This work was supported in part by the National
Security Agency and the Advanced Research and Development Activity
under ARO contract DAAD19-02-1-0079. Partial support in Brazil was
provided by Instituto do Mil\^enio de Nanoci\^encia, CNPq, FAPERJ, and
PRONEX.
\end{acknowledgments}

\appendix
\section{Derivation of Eqs.~(\ref{eq:sol1}), (\ref{eq:sol2}), and (\ref{eq:sol3})}
\label{sec:appendix}

For $t>0$, Eq.~(\ref{eq:hs}) is time-independent: $H_{S} = v_{m}\sigma_{x}$. 
Since the $K$ matrix is also time-independent [Eq.~(\ref{eq:KandPhi})], 
the matrix $\Lambda$ defined by Eq.~(\ref{eq:defoflambda}) is time-independent 
as well. After some straightforward operator algebra, we find that
\begin{eqnarray}
\label{eq:lambdatimeindep}
\Lambda = \frac{1}{2} \int_{0}^{\infty} d\tau \, B(\tau )\, e^{-i\tau
v_{m}\sigma_{x}}\, \sigma_{z}\, e^{i\tau v_{m}\sigma_{x}} \\
\label{eq:lambdatimeindep2}
= \frac{1}{2} \int_{0}^{\infty} d\tau \, B(\tau )
\left[ \sigma_{z} \cos (2v_{m}\tau ) - \sigma_{y} \sin (2v_{m}\tau ) \right].
\end{eqnarray}
One can rewrite Eq.~(\ref{eq:lambdatimeindep2}) as follows
\begin{equation}
\label{eq:lambdagamma}
\Lambda = \frac{1}{2} (\gamma_{1} + i\gamma_{3}) \sigma_{z} -
\frac{1}{2} (\gamma_{2} + i\gamma_{4}) \sigma_{y},
\end{equation}
where $\{\gamma_{i}\}$'s are real coefficients:
%
\begin{equation}
\label{eq:gammas14}
\left\{ 
\begin{tabular}{c}
$\gamma_{1} + i\gamma_{3}$ \\
$\gamma_{2} + i\gamma_{4}$
\end{tabular}
\right\}
= \int_{0}^{\infty} d\tau
\, B(\tau )
\left\{ 
\begin{tabular}{c}
$\cos (2v_{m}\tau )$ \\
$\sin (2v_{m}\tau )$
\end{tabular}
\right\}.
\end{equation}

The density matrix $\rho (t)$ is a $2\times 2$ Hermitian matrix with
unit trace. Hence, it has three real independent components and can be
written as follows:
\begin{equation}
\label{eq:rhoinsigmas}
\rho = \frac{1}{2} + \sigma_{x}\, \mbox{Re}\, \rho_{12} - \sigma_{y}\,
\mbox{Im}\, \rho_{12} + \sigma_{z}\, (\rho_{11} - \frac{1}{2}).
\end{equation}
Let us substitute Eqs.~(\ref{eq:rhoinsigmas}) and
(\ref{eq:lambdagamma}) into the Redfield equation
[Eq.~(\ref{eq:sigmadot})] and use that $H_{S} = v_{m}\sigma_{x}$ and
$K = \frac{1}{2}\sigma_{z}$. A simple algebraic manipulation leads to
three differential equations,
\begin{eqnarray}
\label{eq:de1}
\mbox{Re}\, {\dot \rho}_{12} &=& - \gamma_{1}\, \mbox{Re}\, \rho_{12}
+ \frac{\gamma_{4}}{2},
\\
\label{eq:de2}
{\dot \rho}_{11} &=& -2v_{m}\, \mbox{Im}\, \rho_{12},
\\
\label{eq:de3}
\mbox{Im}\, {\dot \rho}_{12} &=& 
(2v_{m} + \gamma_{2}) (\rho_{11} - \frac{1}{2})
- \gamma_{1}\, \mbox{Im}\, \rho_{12}.
\end{eqnarray}
The initial conditions are $\rho_{11}(0) \!=\! 1$ and $\rho_{12}(0) \!=\! 0$. 
Eq.~(\ref{eq:de1}) decouples from Eqs.~(\ref{eq:de2}) and (\ref{eq:de3}). 
Its solution is given by Eq.~(\ref{eq:sol2}), where we used the following 
identity: $\gamma_{4}/\gamma_{1}=-\tanh (v_{m}/T)$. Eqs.~(\ref{eq:de2}) and
(\ref{eq:de3}) form a closed system. Their solution is given by
Eqs.~(\ref{eq:sol1}) and (\ref{eq:sol3}).

The coefficients $\gamma_{1}$ and $\gamma_{2}$ [Eqs.~(\ref{eq:gamma1})
and (\ref{eq:gamma2}), respectively] are calculated using
Eqs.~(\ref{eq:gammas14}) and (\ref{eq:bath}).




\begin{thebibliography}{30}


\bibitem{nielsen00} M.~A.~Nielsen and I.~L.~Chuang, {\it Quantum
Computation and Quantum Information} (Cambridge University Press,
Cambridge, U.K., 2000).

\bibitem{loss98} D.~Loss and D.~P.~DiVincenzo, Phys. Rev. A {\bf 57},
120 (1998).

\bibitem{divincenzo00} D.~P.~DiVincenzo, D.~Bacon, J.~Kempe,
G.~Burkard, and K.~B.~Whaley, Nature {\bf 408}, 339 (2000).

\bibitem{blick00} R.~H.~Blick and H.~Lorenz, in Proceedings of the
IEEE International Symposium on Circuits and Systems, edited by
J. Calder (IEEE, Piscataway, NJ, 2000), Vol. II, p. 245.

\bibitem{tanamoto00} T.~Tanamoto, 
Phys. Rev. A {\bf 61}, 022305 (2000).

\bibitem{fedichkin04} L.~Fedichkin and A.~Fedorov, Phys. Rev. A {\bf
69}, 032311 (2004); L.~Fedichkin, M.~Yanchenko, and K.~A.~Valiev,
Nanotechnology {\bf 11}, 387 (2000).

\bibitem{brandes02} T.~Brandes and T.~Vorrath, Phys. Rev. B {\bf 66},
075341 (2002).

\bibitem{wu04} Z.-J.~Wu, K.-D.~Zhu, X.-Z.~Yuan, Y.-W.~Jiang, and 
H.~Zheng, preprint cond-mat/0412503.

\bibitem{divincenzo99} D.~P.~DiVincenzo, G.~Burkard, D.~Loss, and
E.~V.~Sukhorukov, in {\it Quantum Mesoscopic Phenomena and Mesoscopic
Devices in Microelectronics}, edited by I.~O.~Kulik and
R.~Ellialtio\u{g}lu (Kluwer, 2000), cond-mat/9911245.

\bibitem{vanderwiel03} W.~G.~van~der~Wiel, S.~De~Franceschi,
J.~M.~Elzerman, T.~Fujisawa, S.~Tarucha, and L.~P.~Kouwenhoven,
Rev. Mod. Phys. {\bf 75}, 1 (2003).

\bibitem{hayashi03} T.~Hayashi, T.~Fujisawa, H.~D.~Cheong,
Y.~H.~Jeong, and Y.~Hirayama, Phys. Rev. Lett. {\bf 91}, 226804
(2003); T.~Fujisawa, T.~Hayashi, H.~D.~Cheong, Y.~H.~Jeong, and
Y.~Hirayama, Physica E {\bf 21}, 1046 (2004).

\bibitem{petta04} J.~R.~Petta, A.~C.~Johnson, C.~M.~Marcus,
M.~P.~Hanson, and A.~C.~Gossard, Phys. Rev. Lett. {\bf 93}, 186802
(2004).

\bibitem{argyres64} P.~N.~Argyres and P.~L.~Kelley, Phys. Rev. {\bf
134}, A98 (1964).

\bibitem{pollard94} W.~T.~Pollard and R.~A.~Friesner,
J. Chem. Phys. {\bf 100}, 5054 (1994).

\bibitem{SpinNote} In the situation we envision, the total number of
electrons would be odd: The ground state of each dot in the absence of
the excess electron would have $S \!=\! 0$, so that the dot with the
excess electron would have spin half. Other situations are, of course,
possible, such as the total number of electrons being even with $S=0$
for both dots when the qubit is in state $\left| 1\right>$ and $S
\!=\! 1/2$ for both in state $\left| 2\right>$; in the latter case,
there would be a singlet and triplet state of the DQD with a small
exchange splitting. Such complications do not effect the underlying
physics that we discuss, and so we neglect them.

\bibitem{brandes99} T.~Brandes and B.~Kramer, Phys. Rev. Lett. {\bf
83}, 3021 (1999).

\bibitem{bruus93} H.~Bruus, K.~Flensberg, and H.~Smith, Phys. Rev. B
{\bf 48}, 11144 (1993).

\bibitem{jeong01} H.~Jeong, A.~M.~Chang, and M.~R.~Melloch, 
Science {\bf 293}, 2221 (2001).

\bibitem{chen04} J.~C.~Chen, A.~M.~Chang, and M.~R.~Melloch, 
Phys. Rev. Lett. {\bf 92}, 176801 (2004).

\bibitem{pulseprofile} The value of $\tau \!\approx\!  15\,{\rm ps}$
is obtained by fitting Eq.~(\ref{eq:biaspulse}) to the experimental
pulse with an effective ramping time of $100\,{\rm ps}$,
Ref.~\onlinecite{fujisawa04}.

\bibitem{itakura03} T.~Itakura and Y.~Tokura, Phys. Rev. B {\bf 67},
195320 (2003).

\bibitem{hartmann04} For another source of decoherence due to coupling
to the leads, see U.~Hartmann and F.~K.~Wilhelm, Phys. Rev. B {\bf
69}, 161309(R) (2004).

\bibitem{fujisawa04} T.~Fujisawa, T.~Hayashi, and Y.~Hirayama, 
J. Vac. Sci. Technol. B {\bf 22}, 2035 (2004).

\bibitem{nazarov93} Yu.~V.~Nazarov, Physica B {\bf 189}, 57 (1993).

\bibitem{gurvitz96} S.~A.~Gurvitz and Ya.~S.~Prager, Phys. Rev. B {\bf
53}, 15932 (1996).

\bibitem{vorojtsov04} S.~Vorojtsov, E.~R.~Mucciolo, and
H.~U.~Baranger, Phys. Rev. B {\bf 69}, 115329 (2004).


\end{thebibliography}
\end{document}